# Kilometer-Scale Neutrino Detectors: First Light

Francis Halzen[a]

[a]Department of Physics, University of Wisconsin-Madison, Madison, WI 53706  USA

**Abstract.** This is a brief report on the status of neutrino "astronomy" at a time when the kilometer-scale neutrino detector IceCube is approaching completion. We revisit the rationale for constructing gigantic neutrino detectors by transforming large volumes of natural ice and water into Cherenkov detectors. With time, the motivation for building such instruments has come into clear focus, and the requirement for their kilometer scale has been rationalized with improved accuracy. We will discuss the performance and some selected results of IceCube based on data taken during construction.

## THE FIRST KILOMETER-SCALE, HIGH-ENERGY NEUTRINO DETECTOR:  ICECUBE

A series of first-generation experiments[1] have demonstrated that high-energy neutrinos with 10 GeV energy and above can be detected by measuring the Cherenkov radiation from secondary particles produced in neutrino interactions inside large volumes of highly transparent ice or water instrumented with a lattice of photomultiplier tubes. The first second-generation detector, IceCube, is under construction at the geographic South Pole[2].

IceCube will consist of 80 km-length strings, each instrumented with 60 10-inch photomultipliers spaced 17 meters apart; see Fig. 1. The deepest module is located at a depth of 2450 meters, so that the instrument is shielded from the large background of cosmic rays at the surface by approximately 1.5 kms of ice. Strings are arranged at apexes of equilateral triangles that are 125 meters on a side. The instrumented detector volume is a cubic kilometer of dark, highly transparent and sterile Antarctic ice. The radioactive background is dominated by instrumentation deployed into this natural ice.

Each optical sensor consists of a glass sphere containing the photomultiplier and the electronics board that digitizes the signals locally using an on-board computer. The digitized signals are given a global time stamp with residuals accurate to less than 3 ns and are subsequently transmitted to the surface. Processors at the surface continuously collect these time-stamped signals from the optical modules; each functions independently.  The digital messages are sent to a string processor and a global event trigger. They are subsequently sorted into the Cherenkov patterns emitted by secondary muon tracks that reveal the direction of the parent neutrino[3].

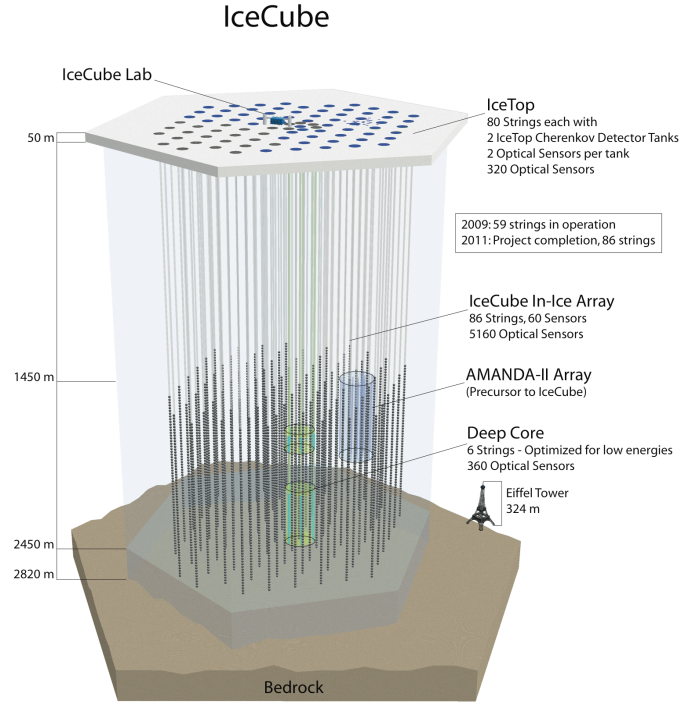

**FIGURE 1.** The IceCube detector, consisting of IceCube, IceTop and the low-energy sub-detector DeepCore. Also shown is the first-generation AMANDA detector.

Based on data taken with 40 of the 59 strings that have already been deployed, the anticipated effective area of the completed IceCube detector is shown in Fig. 2. Notice the factor 2-to-3 increase in effective area over what had been anticipated [4].

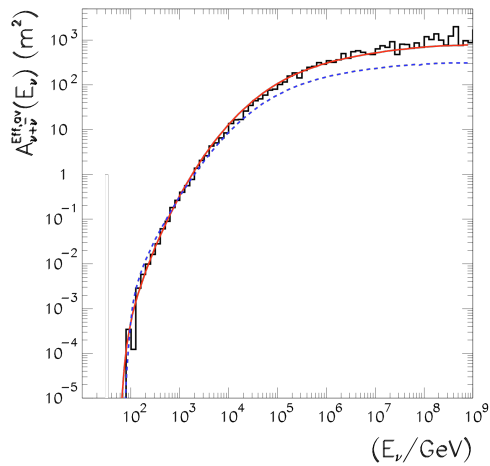

**FIGURE 2.** The neutrino effective area (averaged over the Northern Hemisphere) from IceCube simulation (black histogram) is compared to the convolution of the approximate muon effective area from ref. 5 (solid red line) that we will use in the various estimates of event rates throughout this paper. The neutrino area exceeds the design area (shown as the dashed blue line)[4] at high energy.

The construction of DeepCore, an infill of the IceCube array to be completed by January 2010, lowers the neutrino threshold to 10 GeV neutrino energy over a significant instrumented volume of very clear ice in the lower half of the IceCube detector. It replaces the AMANDA detector, whose operation was terminated in April 2009.

## THE RATIONALE FOR KILOMETER-SCALE NEUTRINO DETECTORS

Despite a discovery potential touching a wide range of scientific issues, construction of IceCube and KM3NeT[6], a concept for a similar detector in the Northern Hemisphere, has been largely motivated by the possibility of opening a new window on the Universe using neutrinos as cosmic messengers. Specifically, we will revisit IceCube's prospects to detect cosmic neutrinos associated with cosmic rays and thus finally reveal their sources prior to the 100th anniversary of their discovery by Victor Hess in 1912.

Cosmic accelerators produce particles with energies in excess of $10^8$ TeV; we still do not know where or how[7]. The flux of cosmic rays observed at Earth is shown in Fig. 3. The energy spectrum follows a sequence of three power laws. The first two are separated by a feature dubbed the "knee" at an energy[a] of approximately 3 PeV. There is evidence that cosmic rays up to this energy are Galactic in origin. Any association with our Galaxy disappears in the vicinity of a second feature in the spectrum referred to as the "ankle"; see Fig. 3. Above the ankle, the gyroradius of a proton in the Galactic magnetic field exceeds the size of the Galaxy, and we are witnessing the onset of an extragalactic component in the spectrum that extends to energies beyond 100 EeV. Direct support for this assumption now comes from two experiments[8] that have observed the telltale structure in the cosmic-ray spectrum resulting from the absorption of the particle flux by the microwave background, the so-called Greissen-Zatsepin-Kuzmin (GZK) cutoff. The origin of the flux in the intermediate region covering PeV-to-EeV energies remains a mystery, although it is routinely assumed that it results from some high-energy extension of the reach of Galactic accelerators.

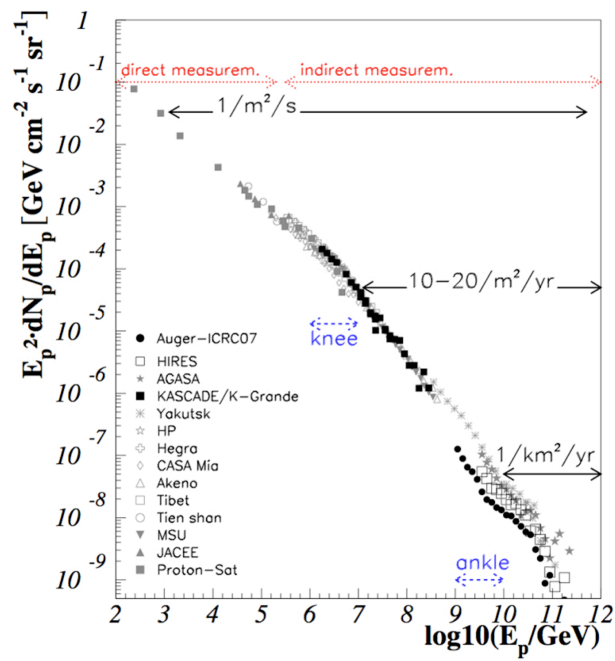

**FIGURE 3.** At the energies of interest here, the cosmic-ray spectrum follows a sequence of 3 power laws. The first 2 are separated by the "knee", the 2nd and 3rd by the "ankle". Cosmic rays beyond the ankle are a new population of particles produced in extragalactic sources.

Acceleration of protons (or nuclei) to TeV energy and above requires massive bulk flows of relativistic charged particles. These are likely to originate from exceptional gravitational forces in the vicinity of black holes or neutron stars. Gravity powers large currents of charged particles that are the origin of high magnetic fields. These create the opportunity for particle acceleration by shocks. It is a fact that electrons are accelerated to high energy near black holes; astronomers detect them indirectly by their synchrotron radiation. Some must accelerate protons, because we observe them as cosmic rays.

How many gamma rays and neutrinos are produced in association with the cosmic-ray beam?[b]. Generically, a cosmic-ray source should also be a beam dump. Cosmic rays accelerated in regions of high magnetic fields near black holes inevitably interact with radiation surrounding them, e.g., UV photons in active galaxies or MeV photons in gamma-ray–burst (GRB) fireballs. In these interactions, neutral and charged pion secondaries are produced by the processes

$$p + \gamma \rightarrow \Delta^+ \rightarrow \pi^o + p \quad \text{and} \quad p + \gamma \rightarrow \Delta^+ \rightarrow \pi^+ + n. \tag{1}$$

While secondary protons may remain trapped in the high magnetic fields, neutrons and the decay products of neutral and charged pions escape. The energy escaping the source is therefore distributed among cosmic rays, gamma rays and neutrinos produced by the decay of neutrons, neutral pions and charged pions, respectively. In the case of Galactic supernova shocks, cosmic rays interact with gas in the disk, e.g. with dense molecular clouds, producing equal numbers of pions of all three charges in hadronic collisions $p + p \rightarrow n\left[\pi^o + \pi^+ + \pi^-\right] + X$.

Kilometer-scale neutrino detectors have the sensitivity to reveal generic cosmic-ray sources with an energy density in neutrinos comparable to their energy density in cosmic rays[10] and pionic TeV gamma rays[11].

---

[a] We will use energy units TeV, PeV and EeV, increasing by factors of 1000 from GeV energy.
[b] We do not discuss cosmic neutrinos directly produced in the interactions of cosmic rays with microwave photons here; extensions of IceCube have been proposed to detect them[9].

# Sources of Galactic Cosmic Rays

Supernova remnants were proposed as possible sources of Galactic cosmic rays as early as 1934 by Baade and Zwicky[12]; their proposal is a matter of debate after more than 70 years. Galactic cosmic rays reach energies of at least several PeV, the "knee" in the spectrum. Their interactions with Galactic hydrogen in the vicinity of the accelerator should generate gamma rays from decay of secondary pions that reach energies of hundreds of TeV. Such sources should be identifiable by a relatively flat energy spectrum that extends to hundreds of TeV without attenuation; they have been dubbed PeVatrons. Straightforward energetics arguments are sufficient to conclude that present air Cherenkov telescopes should have the sensitivity necessary to detect TeV photons from PeVatrons[5].

They may have been revealed by an all-sky survey in ~10 TeV gamma rays with the Milagro detector[13]. Sources are identified that have been located within nearby star-forming regions in Cygnus and in the vicinity of Galactic latitude $l = 40$ degrees; some are not readily associated with known supernova remnants or with non-thermal sources observed at other wavelengths. Subsequently, directional air Cherenkov telescopes were pointed at three of the sources, identifying them as PeVatron candidates with an approximate $E^{-2}$ energy spectrum that extends to tens of TeV without evidence for a cutoff [14,15], in contrast with the best-studied supernova remnants RX J1713-3946 and RX J0852.0-4622 (Vela Junior). It remains to be seen, of course, whether any Milagro sources really do reach photon energies of hundreds of TeV.

Some Milagro sources may actually be molecular clouds illuminated by the cosmic-ray beam accelerated in young remnants located within ~100 pc. One expects indeed that multi-PeV cosmic rays are accelerated only over a short time period, when the remnant transitions from free expansion to the beginning of the Sedov phase and the shock velocity is high. The high-energy particles can produce photons and neutrinos over much longer periods when they diffuse through the interstellar medium to interact with nearby molecular clouds[16]. An association of molecular clouds and supernova remnants is expected in star-forming regions.

Despite the rapid development of instruments with improved sensitivity, it has been impossible to conclusively pinpoint supernova remnants as sources of cosmic rays by identifying gamma rays of pion origin. Eliminating the possibility of a purely electromagnetic origin of TeV gamma rays is challenging. Detecting the accompanying neutrinos would provide incontrovertible evidence for cosmic-ray acceleration in the sources. Particle physics dictates the relation between gamma rays and neutrinos, and basically predicts the production of a $\nu_\mu + \bar{\nu}_\mu$ pair for every two gamma rays seen by Milagro. We conclude that, within uncertainties in the source parameters and detector performance, confirmation that Milagro mapped sources of Galactic cosmic rays should emerge after operating the complete IceCube detector for several years; see Fig. 4.

The quantitative statistics can be summarized as follows. For average values of the parameters describing the flux, the completed IceCube detector could confirm sources in the Milagro sky map as sites of cosmic-ray acceleration at the $3\sigma$ level in less than one year and at the $5\sigma$ level in three years[5]. These results agree with previous estimates[17]. There are intrinsic ambiguities in this estimate. With the performance of IceCube now well understood, these are of astrophysical origin.

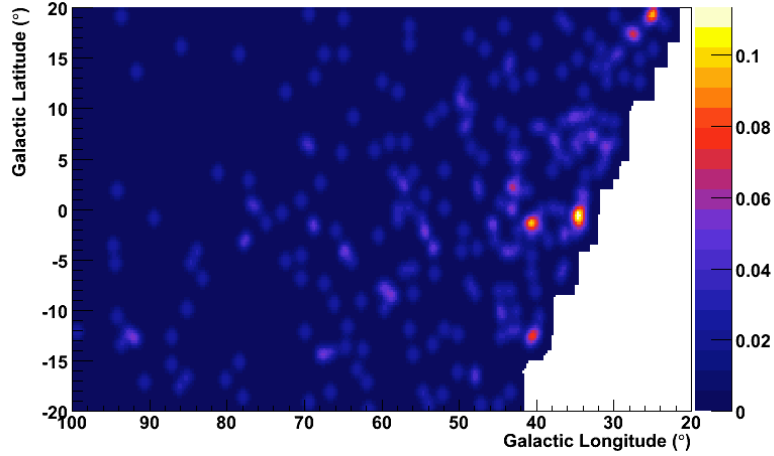

**FIGURE 4.** Simulated sky map of IceCube in Galactic coordinates after 5 years of operation of the completed detector. Two Milagro sources are visible "by eye" with 4 events for MGRO J1852+01 and 3 for MGRO J1908+06 with energy in excess of 40 TeV. These, along with background events, have been randomly distributed according to the resolution of the detector and the size of the sources.

## Sources of Extragalactic Cosmic Rays

Although there is no direct evidence that supernovae accelerate cosmic rays, the idea is generally accepted because of energetics: three supernovae per century converting a reasonable fraction of a solar mass into particle acceleration can accommodate the steady flux of cosmic rays in the Galaxy. Energetics also drive speculation on the origin of extragalactic cosmic rays[10].

By integrating the cosmic-ray spectrum in Fig. 3 above the ankle, we find that the energy density of extragalactic cosmic rays in the Universe is $\rho_E \simeq 3 \times 10^{-19}$ erg cm$^{-3}$. The power required for a population of sources to generate this energy density over the Hubble time of $10^{10}$ years is $3 \times 10^{37}$ erg s$^{-1}$ (Mpc)$^3$. The fireball producing a GRB converts a fraction of a solar mass into the acceleration of electrons, seen as synchrotron photons. The energy in extragalactic cosmic rays can be accommodated with the reasonable assumption that shocks in the expanding GRB fireball convert roughly the same energy into the acceleration of protons that are observed as cosmic rays[18]. It so happens that $2 \times 10^{51}$ erg per GRB will yield the observed energy density in cosmic rays after $10^{10}$ years, given that their rate is 300 per (Gpc)$^3$ per year. Therefore 300 GRBs per year over Hubble time produce the observed cosmic-ray energy density in the Universe, just as three supernovae per century accommodate the steady flux of cosmic rays in the Galaxy.

Cosmic rays and synchrotron photons coexist in the expanding GRB fireball prior to it reaching transparency and producing the observed GRB display. Their interactions produce charged and neutral pions

$$p + \gamma \rightarrow \Delta^+ \rightarrow \pi^o + p \quad \text{and} \quad p + \gamma \rightarrow \Delta^+ \rightarrow \pi^+ + n. \qquad (2)$$

with probabilities 2/3 and 1/3, respectively. Subsequently the pions decay into gamma rays and neutrinos that carry 1/2 and 1/4 of the energy of the parent pion. We here assume that the four leptons in the decay $\pi^+ \rightarrow \nu_\mu + (e + \bar{\nu}_e + \nu_\mu)$ equally share the charged pion's energy. The energy of the pionic leptons relative to the proton is

$$x_\nu = \frac{E_\nu}{E_p} = \frac{1}{2} \langle x_{p \rightarrow \pi} \rangle \simeq \frac{1}{20}, \quad (3)$$

and

$$x_\gamma = \frac{E_\gamma}{E_p} = \frac{1}{2} \langle x_{p \rightarrow \pi} \rangle \simeq \frac{1}{10}. \quad (4)$$

Here

$$\langle x_{p \rightarrow \pi} \rangle = \left\langle \frac{E_\pi}{E_p} \right\rangle \simeq 0.2 \quad (5)$$

is the average energy transferred from the proton to the pion. The secondary neutrino and photon fluxes produced by a GRB are

$$\frac{dN_\nu}{dE} = 1 \, \frac{1}{3} \, \frac{1}{x_\nu} \, \frac{dN_p}{dE_p} \left( \frac{E_p}{x_\nu} \right), \quad (6)$$

$$\frac{dN_\gamma}{dE} = 2 \, \frac{2}{3} \, \frac{1}{x_\gamma} \, \frac{dN_p}{dE_p} \left( \frac{E_p}{x_\gamma} \right) = \frac{1}{8} \, \frac{dN_\nu}{dE}. \quad (7)$$

Here $N_\nu (= N_{\nu_\mu} = N_{\nu_e} = N_{\nu_\tau})$ represents the sum of the neutrino and antineutrino fluxes which are not distinguished by the experiments. Oscillations over cosmic baselines yield approximately equal fluxes for the three flavors.

Neutrinos reach us from sources distributed over all redshifts, while cosmic rays do so only from local sources inside the so-called GZK radius of less than 100 Mpc. The evolution of the sources will boost the neutrino flux by a factor $f_{GZK}$ that depends on the poorly known redshift dependence of GRBs. Assuming a dependence not very different from galaxies, this factor is approximately 3, therefore

$$\frac{dN_\nu}{dE_\nu} = \left[1 - \left(1 - e^{-n_{int}}\right)\right] \frac{1}{3} x_\nu \frac{dN_p}{dE_p} \left( \frac{E_p}{x_\nu} \right) f_{GZK} \simeq n_{int} \, x_\nu \frac{dN_p}{dE_p} \left( \frac{E_p}{x_\nu} \right), \quad (8)$$

where $n_{int}$ is the number of interactions of the proton before escaping the fireball; it is determined by the optical depth of the source for pγ interactions. The key to the production of neutrinos is the fact that protons coexist with GRB gamma rays when the fireball is opaque to gamma-gamma and p-gamma interactions. Fireball phenomenology predicts that, on average, $n_{int} \simeq 1$.

The rate of neutrinos actually detected from this flux can be approximately calculated in the usual way[5]

$$N = 2\pi \times \text{area} \times \text{time} \times \int \frac{dN_\nu}{dE} \, P_{\nu \to \mu} \, dE, \qquad (9)$$

with

$$P_{\nu \to \mu} \simeq 10^{-6} E_\nu (\text{TeV}), \qquad (10)$$

and

$$\frac{dN_\nu}{dE} \simeq x_\nu \frac{5 \times 10^{-11}}{E} \text{ TeV cm}^{-2} \text{ s}^{-1} \text{ sr}^{-1}. \qquad (11)$$

Here we approximated the energy in the cosmic-ray flux to 1 particle per km² per year at 10 EeV. We obtain

$$N \simeq 4 \log\left(\frac{E_{\nu_{max}}}{E_{\nu_{min}}}\right) \simeq 15 \text{ events per year}^{-1}. \qquad (12)$$

The value of the log describing the energy reach of the accelerators is not important, as it cancels the same log that appears in the integration that yields $\rho_E$. The number of events is indeed 15 after cancellation of the two logs. The calculation can be repeated for arbitrary power-law spectra $E^{-(\alpha+1)}$. The number of events is smaller and approaches 15 per kilometer square year when α approaches 1. This calculation neglects the partial absorption of neutrinos in the Earth[19], but this is compensated by the fact that the IceCube effective area exceeds 1 km², as previously discussed.

Problem solved? Not really: the energy density of the extragalactic cosmic rays can be accommodated by active Galactic nuclei provided each converts 2 x 10⁴⁴ ergs⁻¹ into particle acceleration. As is the case for GRBs, this is an amount that matches their output in electromagnetic radiation. The calculation just performed for GRBs can be repeated assuming that active Galactic nuclei (AGN) are the sources of the cosmic rays. In this case, the site of cosmic-ray production is a matter of speculation, and the nature of the target photons is uncertain. It is therefore difficult to estimate $n'_\gamma$ [20].

Whether GRBs or AGN, the observation that these sources radiate similar energies in photons and cosmic rays is consistent with the beam-dump scenario previously discussed. In the interaction of cosmic rays with radiation and gases near a black hole, roughly equal energy goes into the secondary protons (or neutrons) and neutral pions whose energy ends up in cosmic rays and gamma rays, respectively. It predicts a matching flux of neutrinos that, after many correction factors that cancel, is roughly equal to, or a fraction of, the cosmic-ray flux previously introduced:

$$E_\nu^2 \frac{dN}{dE_\nu} = 5 \times 10^{-11} \text{ TeV cm}^{-2} \text{ s}^{-1} \text{ sr}^{-1} \qquad (13)$$

If we adjust it downward by a factor $x_\nu$, we obtain a generic neutrino flux predicted by the GRB and AGN scenarios. After seven years of operation, AMANDA's sensitivity is approaching the interesting range, but it takes IceCube to explore it; see Fig. 5.

If it turns out that GRBs are the sources, IceCube's mission is relatively straightforward, because we expect to observe of order 10 neutrinos per kilometer square per year[19] in coincidence with Swift and Fermi GRBs, which translates into a 5σ observation. A similar statistical power can be obtained by detecting showers produced by electron and tau neutrinos.

In summary, while the road to identification of sources of the Galactic cosmic ray has been mapped, the origin of the extragalactic component remains unresolved.

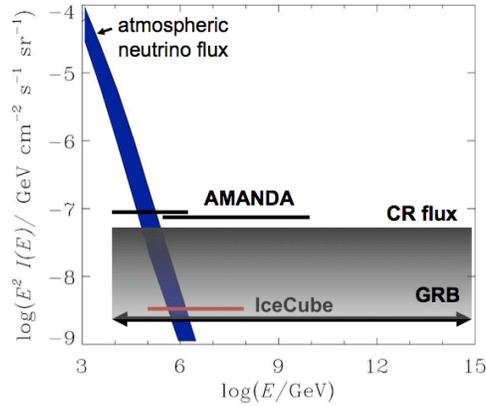

**FIGURE 5.** Our energetics estimate of the flux of neutrinos associated with sources of the highest-energy cosmic rays (the shaded region in the plot ranges from the cosmic-ray flux to the flux derived assuming that GRBs are the sources of extragalactic cosmic rays) is compared to upper limits established by AMANDA and the sensitivity of IceCube[21]. Also shown is the background flux of atmospheric neutrinos at lower energies.

## ICECUBE: FIRST LIGHT

The present status[22] of the search for point sources of cosmic neutrinos is shown in Fig. 6. The neutrino map is the result of an unbinned search, a method[23] that improved the sensitivity of IceCube by a factor of two over the conventional binned methods previously used[24]. Potential sources of neutrinos are not only identified by clustering in arrival direction; the analysis also takes into account that their energy may be large and less likely to be accommodated by the atmospheric background of relatively low-energy events. Note that, with this method, IceCube has sensitivity to high-energy sources over the full sky. With one half-year of data taken by a detector instrumenting one-half kilometer-cubed of ice, we are not yet sensitive to the predictions for neutrinos associated with cosmic-ray sources previously discussed. With a growing detector, we expect to reach the required neutrino exposure within a year or two.

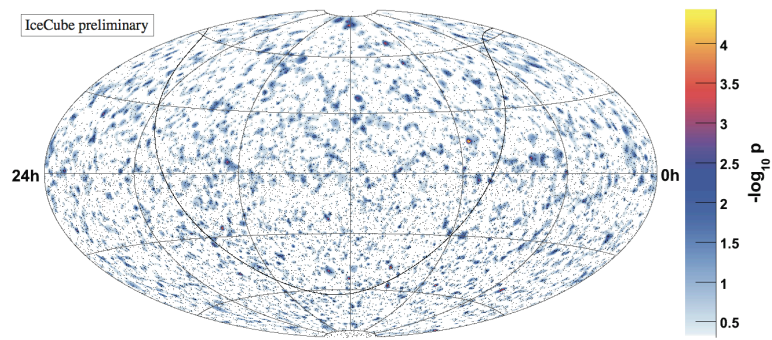

**FIGURE 6.** Using declination and right ascension as coordinates, the map shows the probability for a point source of high-energy neutrinos with energies not readily accommodated by the steeply-falling atmospheric neutrino flux. The map was obtained by operating IceCube with 40 strings for half a year. This is the first result obtained with half of IceCube instrumented. The "hottest spot" in the map represents an excess of 7 events, an excursion from the atmospheric background with a probability of $10^{-4.4}$. After taking into account trial factors, the probability for this event to happen anywhere in the sky map is not significant. The map consists of 6796 neutrinos in the Northern Hemisphere and 10,981 down-going muons rejected to the $10^{-5}$ level in the Southern Hemisphere, shown as black dots.

IceCube may very well discover the particle nature of dark matter[25]. The detector searches for neutrinos produced by the annihilation of dark-matter particles gravitationally trapped at the center of the Sun and the Earth. In searching for generic weakly interacting, massive dark-matter particles (WIMPs) with spin-independent interactions with ordinary matter, IceCube is only competitive with direct-detection experiments if the WIMP mass is sufficiently large. On the other hand, for WIMPs with mostly spin-dependent interactions, IceCube[26] and AMANDA[27] have already improved on the previous best limits obtained by the SuperK experiment using the same method. They improve on the best limits from direct-detection experiments by two orders of magnitude; see Fig. 7. The AMANDA limit extending to lower WIMP masses resulted from an analysis of AMANDA data obtained after seven years of operation. It rules out supersymmetric WIMP models not excluded by other experiments. Using the same data set, we measured the atmospheric neutrino flux up to 100-TeV energy. The agreement with expectations resulted in best limits on new physics, such as violations of Lorentz invariance and of the equivalence principle or non-standard neutrino interactions[28].

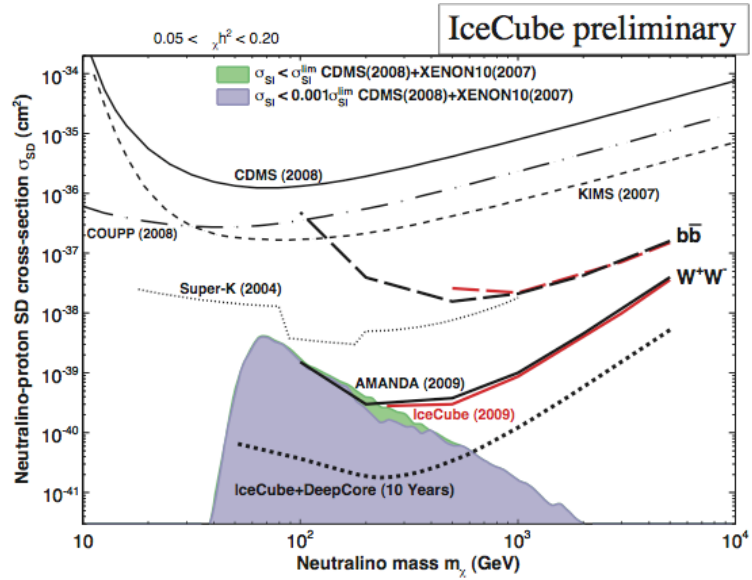

**FIGURE 7.** The solid and dashed red lines show the upper limits at 90% confidence level on the spin-dependent interaction of dark matter particles with ordinary matter. The two lines represent extreme cases where neutrinos originate mostly from heavy quarks (top line) and weak bosons (bottom line) produced in the annihilation of dark-matter particles. Also shown is the reach of the complete IceCube and its DeepCore extension after 5 years of observation of the Sun. The shaded area represents supersymmetric models not disfavored by direct searches for dark matter. Also shown are previous limits from direct experiments and from the SuperK experiment. The results are noteworthy in that they improve by 2 orders of magnitude on the sensitivity previously obtained by direct experiments.

IceCube is a huge cosmic-ray muon detector and the first sizeable detector covering the Southern Hemisphere. Using samples of several billion downward-going muons, we are studying enigmatic anisotropies recently identified in the cosmic-ray spectrum by Northern detectors; see Fig. 8. IceCube data show that these anisotropies persist at energies in excess of 100 TeV, ruling out the Sun as their origin[29]. Where the interpretations are concerned, we are investigating possibilities that the asymmetry in arrival directions of cosmic rays are associated with structure in the Galactic magnetic field, or with diffusive particle flows from nearby Galactic sources such as Vela.

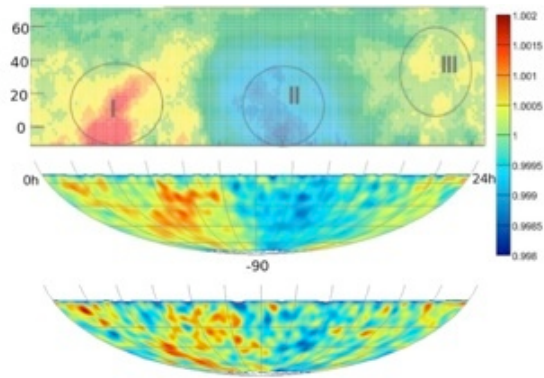

**FIGURE 8.** The anisotropy in arrival directions of cosmic rays with an average energy of 10 TeV is plotted in equatorial coordinates[28]. Top: Tibet Array observation in the Northern Hemisphere.[30] Region I represents the excess in cosmic-ray intensity, while region II shows the deficit, and region III shows an enhanced flux toward the Cygnus region. Middle: arrival direction of cosmic-ray muons detected with 22 IceCube strings. The color scale represents the relative intensity. Note that the two observations match, giving a clear full-sky view. The origin of this effect is unknown. Bottom: While the upper map displays the arrival directions of all muons with an average energy of 12 TeV, the lower map displays the arrival directions of muons with an average energy of 125 TeV.

## ACKNOWLEDGMENTS

I would like to thank my collaborators Concha González-García, Darren Grant, Alexander Kappes and Aongus O'Murchadha, as well as John Beacom, Julia Becker, Peter Biermann, Martino Olivo, Christian Spiering and Stefan Westerhoff for valuable discussions. This research was supported in part by the U.S. National Science Foundation under Grants No. OPP-0236449 and PHY-0354776; by the University of Wisconsin Research Committee with funds granted by the Wisconsin Alumni Research Foundation; and by the Alexander von Humboldt Foundation in Germany.

# REFERENCES

1. C. Spiering, "High Energy Neutrino Astronomy: Status and Perspectives," presented at the Intl. Symposium on High-Energy Gamma-Ray Astronomy, Heidelberg, 2008; astro-ph/0811.4747.
2. IceCube collaboration, *Astropart. Phys.* **26** 155 (2006), astro-ph/0604450; T. Montaruli f. t. IceCube collaboration, "First Results of the IceCube Observatory on High-Energy Neutrino Astronomy" in *Proc. of TAUP 07*, edited by K. Inoue, *et al.*, Jour. Phys. Conf. Series **120** 022009 (2008), astro-ph/0712.3524; and S.R. Klein f. t. IceCube collaboration, "IceCube: A Cubic-Kilometer Radiation Detector" in *Proc. of SORMA West 2008* IEEE Trans. Nucl. Sci. **56** 1141 (2009), physics.ins-det/0807.0034.
3. F. Halzen, Eur. Phys. J. C **46** 669 (2006); astro-ph/0602132.
4. IceCube collaboration, *Astropart. Phys.* **20** 507 (2004), astro-ph/0305196.
5. See M.C. González-García, *et al.*, Astropart. Phys. **31** 6 437 (2009); astro-ph.HE/0902.1176, and references therein.
6. E. Migneco, J. Phys. Conf. Ser. **136** 022048 (2008).
7. For recent reviews of the cosmic-ray problem, see P. Sommers and S. Westerhoff, astro-ph/0802.1267; A.M. Hillas, astro-ph/0607109; and V. Berezinsky, astro-ph/0801.3028.
8. HiRes collaboration, Phys. Rev. Lett. **100** 101101 (2008), astro-ph/0703099; and Pierre Auger collaboration, Phys. Rev. Lett. **101** 061101 (2008), astro-ph/0806.4302.
9. P. Allison, *et al.*, Nucl. Instr. and Meth. A **604** S64 (2009); astro-ph.HE/0904.1309.
10. T.K. Gaisser, OECD Megascience Forum (1997), astro-ph/9707283; the discussion was recently revisited in Ahlers, *et al.*, Phys. Rev. D **72** 023001 (2005), astro-ph/0503229.
11. J. Alvarez-Muñiz and F. Halzen, Astrophys. J. **576** L33 (2002); astro-ph/0205408.
12. W. Baade and F. Zwicky, Phys. Rev. D **46** 76 (1934).
13. A.A. Abdo, *et al.*, Astrophys. J. **658** L33 (2007); astro-ph/0611691.
14. H.E.S.S. collaboration, "Galactic Plane Survey Unveils a Milagro Hotspot" in Proc. of 30[th] ICRC **2** 863 OG2.2 (2007); astro-ph/07102418.
15. MAGIC collaboration., Astrophys. J. Lett. **675** L25 (2008); astro-ph/0801.2391.
16. For a recent discussion, see S. Gabici and F.A. Aharonian, Astrophys. J. **665** L131; astro-ph/0705.3011.
17. F. Halzen, *et al.*, Phys. Rev. D **78** 063004 (2008), see also Nucl. Instr. and Meth A **602** 117 (2009); astro-ph/0803.0314v2.
18. E. Waxman, Phys. Rev. Lett. **75** 386 (1995), astro-ph/9701231; M. Vietri, Phys. Rev. Lett. **80** 3690 (1998), astro-ph/9802241; and M. Bottcher and C.D. Dermer, New J. Phys. **8** 122, astro-ph/9801027.
19. D. Guetta, *et al.*, Astropart. Phys. **20** 429 (2004); astro-ph/0302524.
20. F. Halzen and A. O'Murchadha, "Neutrinos from Auger Sources" in *Proc. of 4[th] Intl. Workshop on Neutrino Oscillations in Venice*, edited by M. Baldo Ceolin, 159 (2008); astro-ph/0802.0887.
21. G.C. Hill f. t. IceCube collaboration, "Likelihood Deconvolution of Diffuse Prompt and Extra-Terrestrial Neutrino fluxes in the AMANDA-II Detector" in Proc. of 30[th] ICRC **5** 1453 (2007); astro-ph/0711.0353 79; L. Gerhardt f. t. IceCube collaboration, "Multi-year Search for UHE Diffuse Neutrino Flux with AMANDA-II" in Proc. of 30[th] ICRC **5** 1429 (2007); astro-ph/0711.0353 75.
22. T. Montaruli, "Rapporteur Summary of Sessions HE 2.2-2.4 and OG 2.5-2.7" in Proc. of 31[st] ICRC (2009); hep-ph/0910.4364.
23. J. Braun, *et al.*, Astropart. Phys. **29** 299 (2008); astro-ph/0801.1604.
24. IceCube collaboration, Phys. Rev. Lett. **102** 201302 (2009); astro-ph.HE/0902.2460.
25. F. Halzen and D. Hooper, New J. Phys. **11** 105019 (2009); astro-ph.HE/0910.4513.
26. IceCube collaboration, Phys. Rev. Lett. **102** 201302 (2009); astro-ph.CO/0902.2460.
27. J. Braun and D. Hubert for the IceCube collaboration, "Searches for WIMP Dark Matter from the Sun with AMANDA" in Proc. of 31[st] ICRC HE2.3 (2009); astro-ph.HE/0906.1615.
28. IceCube collaboration, Phys. Rev. D **79** 102005 (2009); astro-ph.HE/0902.0675.
29. R. Abbasi and P. Desiati for the IceCube collaboration, "Large-scale Cosmic-ray Anisotropy with IceCube" in Proc. of 31[st] ICRC SH3.2 (2009); astro-ph.HE/0907.0498.
30. The Tibet AS-Gamma collaboration, Astrophys. J. **633** 1005 (2005); astro-ph/050239.